\documentclass[10pt]{article}
\usepackage[a4paper]{geometry}
\usepackage{amssymb}
\usepackage[font=normalsize,format=plain,labelfont=sc]{caption}
\usepackage{graphicx,epsfig}

\begin{document}

\newcommand{\E}{\mbox{$\mathbb{E}$}}
\newcommand{\bm}[1]{\mbox{\boldmath $ #1 $ }}

\newtheorem{definition}{Definition}[section]
\newtheorem{theorem}{Theorem}[section]
\newtheorem{lemma}{Lemma}[section]
\newtheorem{conjecture}{Conjecture}[section]
\newtheorem{observation}{Observation}[section]
\newtheorem{corollary}{Corollary}[section]
\newtheorem{claim}{Claim}[section]
\newtheorem{fact}{Fact}[section]
\newtheorem{proposition}{Proposition}[section]

\newenvironment{proof}{\begin{trivlist}
                       \item[]\hspace{0cm}{\bf Proof}
                       \hspace{0cm} }{\hfill $\Box$
                       \end{trivlist}}

\newenvironment{sketchproof}{\begin{trivlist}
                       \item[]\hspace{0cm}{\bf Sketch of Proof}
                       \hspace{0cm} }{\hfill $\Box$
                       \end{trivlist}}

\newenvironment{remark}{\begin{trivlist}
                       \item[]\hspace{0cm}{\bf Remark}}{
                       \end{trivlist}}

\newenvironment{remarks}{{\bf Remarks}{\flushleft}
                         \begin{list}{-}{\setlength{\itemsep}{0.5\baselineskip}
                                         \setlength{\topsep}{0.5\baselineskip}
                                         \setlength{\parsep}{0cm}
                                         \setlength{\parskip}{0cm}}}{
                                         \vspace{0.5\baselineskip}
                          \end{list}}

\bibliographystyle{article}
\def\bibfmta#1#2#3#4{#1, {#2}, {\sl #3}, #4.}
\def\bibfmtb#1#2#3#4{{#1}, {\sl #2}, { #3}, #4.}
\def\bibfmtib#1#2#3 {{#1}, {\bf #2}, {\em #3}}

\title{\bf Analysis of an Efficient Distributed \\
Algorithm for Mutual Exclusion \\
(Average-Case Analysis of Path Reversal)}

\author{Christian LAVAULT}
\date{INSA/IRISA (CNRS URA 227)\\
20 Avenue des Buttes de Co\"{e}smes 35043 Rennes Cedex, France.\\
\mbox{Email}~: lavault@irisa.fr
}

\maketitle

\begin{abstract}
The algorithm designed in \cite{nta,tn} was the very first distributed
algorithm to solve the mutual exclusion problem in complete networks by
using a dynamic logical tree structure as its basic distributed data
structure, {\em viz.} a {\em path reversal transformation} in rooted
$n$-node trees; besides, it was also the first one to achieve a logarithmic 
average-case message complexity. The present paper proposes a direct and 
general approach to compute the {\em moments of the cost of path reversal}. 
It basically uses one-one correspondences between combinatorial structures 
and the associated probability generating functions: the expected cost of 
path reversal is thus proved to be exactly $H_{n-1}$.
Moreover, time and message complexity of the algorithm as well as randomized
bounds on its worst-case message complexity in arbitrary networks are also
given. The average-case analysis of path reversal and the analysis of this 
distributed algorithm for mutual exclusion are thus fully completed in the paper. 
The general techniques used should also prove available and fruitful when adapted
to the most efficient recent tree-based distributed algorithms for mutual
exclusion which require powerful tools, particularly for average-case analyses.
\end{abstract}

\section{Introduction}
A distributed system consists of a collection of geographically dispersed
autonomous sites, which are connected by a communication network. The
sites (or processes) have no shared memory and can only communicate
with one another by means of messages. \par
In the {\em mutual exclusion problem}, concurrent access to a shared
resource, called the {\em critical section} ({\em CS}), must be synchronized
such that at any time, only one process can access the ({\em CS}). Mutual exclusion
is crucial for the design of distributed systems. Many problems involving
replicated data, atomic commitment, synchronization, and others require that
a resource be allocated to a single process at a time. Solutions to this
problem often entail high communication costs and are vulnerable to site and
communication failures.
\par
Several distributed algorithms exist to implement mutual exclusion
\cite{ab,ca,la,ma,ra,ri,tn}, etc., they usually are designed for complete or
general networks and the most recent ones are often fault tolerant. But,
whatever the algorithm, it is either a permission-based, or a token-based
algorithm, and thus, it uses appropriate data structures. Lamport's
token-based algorithm \cite{la} maintains a waiting queue at each site and
the message complexity of the algorithm is $3(n - 1)$, where $n$ is the
number of sites.
Several algorithms were presented later, which reduce the number of messages
to $\Theta(n)$ with a smaller constant factor \cite{ca,ri}. Maekawa's
permission-based algorithm \cite{ma} imposes a logical structure on the
network and only requires $c\sqrt{n}$ messages to be exchanged (where $c$
is a constant which varies between 3 and 5).
\par
The token-based algorithm $\cal A$ (see \cite{nta,tn}), which is analysed
in the present paper, is the first mutual exclusion algorithm for complete
networks which achieves a logarithmic
average message complexity~; besides, it is the very first one to use a {\em
tree-based} structure, namely a path reversal, as its basic distributed data
structure. More recently, various mutual
exclusion algorithms ({\em e.g.} \cite{ab,ra}, etc.) have been designed
which use either the same data structure, or some very close tree-based data
structures. They usually also provide efficient (possibly fault
tolerant) solutions to the mutual exclusion problem.
\par
The general model used in \cite{nta,tn} to design algorithms $\cal A$
assumes the underlying communication links and the processes to be
reliable. Message propagation delay is finite but impredictable and the
messages are not assumed to obey the FIFO rule. A process entering the
({\em CS}) releases it within a
finite delay. Moreover, the communication network is {\em complete}. To
ensure a fair mutual exclusion, each node in the network maintains two
pointers, {\em Last} and {\em Next}, at any time. {\em Last} indicates the
node to which requests for ({\em CS}) access should be forwarded~; {\em
Next} points to the node to which  access permission must be forwarded after
the current node has executed its own ({\em CS}). As described below, the
dynamic updating of these two pointers involves two distributed data
structures:
a waiting queue, and a {\em dynamic} logical rooted tree structure which is
nothing but a path reversal. Algorithm $\cal A$ is thus very
efficient in terms of  average-case message
complexity, {\em viz.} $H_{n-1} = \ln n + O(1)$\footnote{Throughout the
paper, $\lg$
denotes the base two logarithm and $\ln$ the natural logarithm. 
$H_{n} =\: \sum_{i=1}^{n} 1/i$ denotes the $n$-th harmonic number,
with asymptotic value $H_{n} = \:\ln n \: + \: \gamma \:+\: 1/2n \:+\: O(n^{-2})$ 
(where $\gamma = 0.577\ldots$ is Euler's constant)}.
\par
Let us recall now how the two data
structures at hand are actually involved in the algorithm, which is fully
designed in \cite{nta,tn}.
Algorithm $\cal A$ uses the notion of {\em token}. A node
can enter its ({\em CS}) only if it has the token. However, unlike the
concept of a token circulating continuously in the system, the token is sent
from one node to another if and only if a request is made for it. The token
(also called {\em privilege message}) consists of a  queue of processes which
are requesting the ({\em CS}). The token circulates strictly according
to the order in which the requests have been made.
\par
The first data structure
used in $\cal A$ is a {\em waiting queue} which is updated by
each node after executing its own ({\em CS}). The waiting queue of
requesting processes is maintained at the
node containing the token and is transferred along with the token whenever
the token is transferred. The requesting nodes receive the token strictly
according to the order in the queue.
Each node knows its next node in the waiting queue only if the {\em Next}
exists. The head is the node which owns the token and the tail is the last
node which requested the ({\em CS}). Thus, a path is constructed in such a
way that each request message is transmitted to the tail. Then, either the
tail is in the ({\em CS}) and it let the requesting node enter it, or the
tail waits for the token, in which case the requesting node is appended to
the tail.
\par
The second data structure involved in algorithm $\cal A$ gives the path to
go to the tail: it is a logical rooted ordered tree. A node which requests
the ({\em CS}) sends its message to its {\em Last}, and, from {\em Last} to
{\em Last}, the request is transmitted to the tail of the waiting queue. In
such a structure, every node knows only its {\em Last}. Moreover, if the
requesting node is not the last, the logical tree structure is transformed:
the requesting node is the new {\em Last} and the nodes which are located
between the requesting node and the last will gain the new last as
{\em Last}.
This is typically a logical transformation of {\em path reversal}, which is
performed at a node $x$ of an ordered $n$-node tree $T_{n}$ consisting of
a root with $n - 1$ children. These transformations  $\varphi(T_{n})$
are performed to keep a dynamic decentralized path towards the tail of the
waiting queue.
\par
In \cite{gi}, Ginat, Sleator and Tarjan derived a tight upper bound of
$\lg n$ for the cost of path reversal in using the notion of {\em amortized
cost} of a path
reversal. Actually, by means of combinatorial and algebraic methods on the
Dycklanguage (namely by encoding oriented ordered trees $T_{n}$ with
Dyckwords), the average number of messages used by algorithm
$\cal A$ was obtained in \cite{nta}.
By contrast, the present paper uses direct and general derivation methods
involving one-to-one correspondences
between combinatorial structures such as priority queues, binary tournament
trees and permutations.
Moreover, a full analysis of algorithm $\cal A$ is completed in this paper
from the computation of the first and second moments of the cost of path
reversal~; {\em viz.} we derive the expected and worst-case message
complexity of $\cal A$ as well as its average and
worst-case waiting time.  Note that the average-case analysis of
other efficient mutual exclusion tree-based algorithms ({\em e.g.}
\cite{ab,ra}, among others) may easily be adaptated from the present one,
since the data structures involved in such algorithms are quite close to
those of algorithm $\cal A$.
The analysis of the average waiting time using simple birth-and-death
process methods and asymptotics, it could thus also apply easily to the
waiting time analysis of the above-mentioned algorithms. In this sense, the
analyses proposed in this paper are quite general indeed.

The paper is organized as follows.
In Section 2, we define the path reversal transformation performed in a
tree $T_{n}$ and give a constructive proof of the one-one correspondence
between priority queues and the combinatorial structure of trees $T_{n}$.
In Section 3, probability generating functions are computed which
yield the exact expected cost of path reversal: $H_{n-1}$, and the second
moment of the cost. Section 4 is devoted to the computation of the waiting
time and the expected waiting time of algorithm $\cal A$. In Section 5,
more extended complexity results are given, {\em viz.} randomized bounds
on the worst-case message complexity of the algorithm in {\em arbitrary}
networks. In the Appendix, we propose a second proof technique which directly
yields the exact expected cost of path reversal by solving a straight and
simple recurrent equation.

\section{One-one correspondences between combinatorial structures}
We first define a path reversal transformation in $T_{n}$ and its {\em cost}
(see \cite{gi}). Then we point out some one-to-one correspondences between 
combinatorial objects and structures which are relevant to the problem of computing 
the average cost of path reversal. Such one-to-one tools are used in Section 3
to compute this expected cost and its variance by means of corresponding
probability generating functions.

\subsection{Path reversal}
Let $T_{n}$ be a rooted $n$-node tree, or an ordered tree with $n$ nodes, 
according to either \cite{gi}, or \cite[page 306]{kn}. A {\em path reversal}
at a node $x$ in $T_{n}$ is performed by traversing the path from $x$ to the 
tree root $r$ and making $x$ the parent (or pointer {\em Last}) of each node 
on the path other than $x$. Thus $x$ becomes the new tree root.
The {\em cost} of the reversal is the number of edges on the path
reversed. Path reversal is a variant of the standard path compression
algorithm for maintaining disjoint sets under union.

\begin{figure}[t]
    \center
    \includegraphics[width=0.9\textwidth]{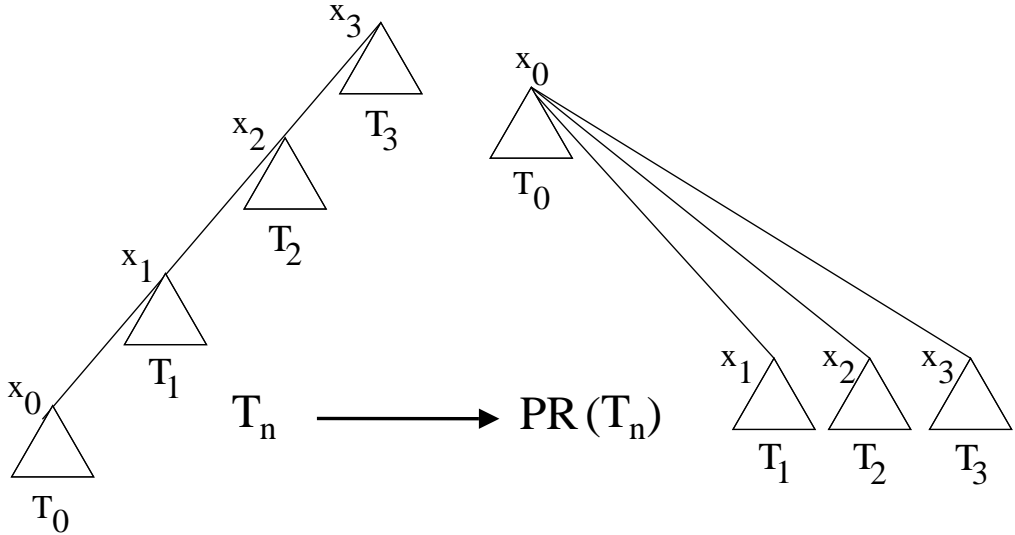}
    \caption{Path reversal $\varphi_{x_{0}}$.
    The $T_{i}$'s denote the (left/right) subtrees of $T_{n}$.}
\end{figure}

The average cost of a path reversal performed on an initial ordered $n$-node
tree $T_{n}$ which consists of a root with $n - 1$ descendants (or
{\em children}, as in \cite{gi}) is the expected number of edges on the paths
reversed in $T_{n}$ (see Figure 1).
In words, it is the {\em expected height of such reversed trees}
$\varphi(T_{n})$, provided that we let the
height of a tree root be 1: {\em viz.} the {\em height} of a node $x$ in
$T_{n}$ is thus defined as being the number of nodes on the path from the
node $x$ to the root $r$ of $T_{n}$.
\par
It turns out that the average number of messages used in $\cal A$ is
actually the expected cost of a path reversal performed on such initial
ordered $n$-node
trees $T_{n}$ which consist of a root with $n - 1$ children. This is indeed
the average number of changes of the variable {\em Last} which builds the
dynamic data structure of path reversal used in algorithm $\cal A$.

\subsection{Priority queues, tournament trees and permutations}
Whenever two combinatorial structures are counted by the same number,
there exist one-one mappings between the two structures.
Explicit one-to-one correspondences between combinatorial representations
provide coding and decoding algorithms between the stuctures. We now need the
following definitions of some combinatorial structures which are closely
connected with path reversal and involved in the computation of its cost.

\subsubsection{Definitions and notations}
\begin{description}
\item
{\sl (i)} Let $[n]$ be the set $\{1, 2, \ldots , n\}$.
A {\em permutation} is a {\em one-one mapping} $\sigma : [n] \rightarrow
[n]$; we write $\sigma \in S_{n}$, where $S_{n}$ is the symmetric group over
$[n]$.
\item
{\sl (ii)} A {\em binary tournament tree} of size $n$ is a binary $n$-node
tree whose internal nodes are labeled with consecutive integers of $[n]$, in
such a way that the root is labeled 1, and all labels are decreasing
({\em bottom-up}) along each branch. Let ${\cal T}_{n}$ denote the set of
all binary tournament trees of size $n$.
${\cal T}_{n}$ also denotes the set of tournament representations of all
permutations $\sigma~\in~S_{n}$, considered as elements of $[n]^{n}$,
since the correspondence $\tau : S_{n} \rightarrow {\cal T}_{n}$ is
one-one (see \cite{vu} for a detailed proof). Note that this one-to-one
mapping implies that $\left| {\cal T}_{n} \right| = n!$
\item
{\sl (iii)} A {\em priority queue} of size $n$ is a set $Q_{n}$ of keys ;
each key $K \in Q_{n}$ has an associated priority $p(K)$ which is an
arbitrary integer. To avoid cumbersome
notations, we identify $Q_{n}$ with the set of priorities of its keys.
Strictly speaking, this is a set with repetitions since priorities need not
be all distincts. However, it is convenient to ignore this technicality
and assume {\em distinct priorities}. The simplest representation of a
priority queue of size $n$ is then a sequence
$s~=~(p_{1},p_{2},~\ldots~,p_{n})$ of the priorities of $Q_{n}$, kept in
their order of arrival. Assume the $n!$ possible orders of arrival of the
$p_{i}$'s
to be equally likely, a priority queue $Q_{n}$ ({\em i.e.} a sequence $s$ of
$p_{i}$'s) is defined as
random {\em iff} it is associated to a random order of the $p_{i}$'s.
There is a one-to-one correspondence between the set ${\cal T}_{n}$ of all the
$n$-node binary tournament trees and the set of all the priority queues
$Q_{n}$ of size $n$.
To each one sequence of priorities $s = (p_{1}, \ldots ,p_{n}) \in Q_{n}$,
we associate a binary tournament tree $\gamma(s) = T \in {\cal T}_{n}$
by the following rules: let {\bf m}$ \; = \; \min(s)$, we then write $s =
\ell \; \mbox{\bf m} \; r$; the binary tree $T \in {\cal T}_{n}$ possesses
{\bf m} as root, $\gamma(\ell)$ as left subtree and $\gamma(r)$ as right
subtree. The rules are applied repeatedly to all the left and right
subsequences of $s$, and from the root of $T$ to the leaves of $T$; by
convention, we let $\gamma(\emptyset) = \Lambda$ (where $\Lambda$
denotes the empty binary tree). The correspondence $\gamma$ is obviously
one-one (see \cite{fr} for a fully detailed constructive proof).
\par
We shall thus use binary tournaments ${\cal T}_{n}$ to represent the
permutations of $S_{n}$ as well as the priority queues $Q_{n}$ of size $n$.
\item
{\sl (iv)} If $T \in {\cal T}_{n}$ is a binary tournament, its
{\em right branch} $RB (T)$ is the increasing sequence of
priorities found on the path starting at the root of $T$ and repeatedly
going to the right subtree. The {\em bottom} of $RB (T)$ is the node
having no right son. The {\em left branch} $LB (T)$ of $T$ is defined in a
symmetrical manner.
\end{description}

\subsection{The one-one correspondence between \protect\bm{Q_{n}} and
\protect\bm{T_{n}}}
We now give a constuctive proof of a {\em one-to-one correspondence} mapping
the given combinatorial structure of ordered trees $T_{n}$ (as defined in
the Introduction) onto the priority queues $Q_{n}$.
\begin{theorem}
There is a one-to-one correspondence between the priority queues of size
$n, Q_{n}$, and the ordered $n$-node trees $T_{n}$ which consist of a root with
$n - 1$ children.
\end{theorem}
\begin{proof}
There are many representations of priority queues $Q_{n}$ ; let us
consider the $n$-node {\em binary heap} structure, which is very simple and
perfectly suitable for the constructive proof.
\begin{itemize}
\item First, a {\em binary heap} of size $n$ is an {\em essentially complete
binary tree}. A binary tree is {\em essentially complete} if each of its
internal nodes possesses exactly two children, with the possible exception
of a unique {\em special} node situated on level $(h - 1)$ (where $h$ denotes 
the height of the heap), which may possess only a left child and no right child.
Moreover, all the leaves are either on level $h$, or else they are on levels
$h$ and $(h-1)$, and no leaf is found on level $(h-1)$ to the left of an
internal node at the same level. The unique special node, if it exists, is
to the right of all the other level $(h-1)$ internal nodes in the subtree.\\
Besides, each tree node in a binary heap contains one item, with the items
arranged in heap order ({\em i.e.} the priority queue ordering): the key of
the item in the parent node is strictly smaller than the key of the item in
any descendant's node.
Thus the root is located at position 1 and contains an item of minimum key.
If we number the nodes of such a essentially complete binary tree from 1 to
$n$ in heap order and identify nodes with numbers, the parent of the node located 
at position $x$ is located at $\lfloor x/2 \rfloor$. Similarly, The left 
son of node $x$ is located at $2x$ and its right son at $\min\{2x + 1,n\}$.
We can thus represent each node by an integer and the entire binary heap
by a map from $[n]$ onto the items: the binary heap with $n$ nodes fits well 
into locations $1, \ldots , n$. This forces a breadth-first, left-to-right 
filling of the binary tree, {\em i.e.} a heap or priority queue ordering.

\item Next, it is well-known that any ordered tree with $n$ nodes may easily 
be transformed into a binary tree by the {\em natural correspondence} between
ordered trees and binary trees. The corresponding binary tree is obtained by
linking together the brothering nodes of the given ordered tree and removing
vertical links except from a father to its first (left) son. \\
Conversely, it is easy to see that any binary tree may be represented as an 
ordered tree by reversing the process. The correspondence is thus one-one 
(see \cite[Vol.~1, page 333]{kn}).
\end{itemize}
Note that the construction of a binary heap of size $n$ can be carried out
in a linear time, and more precisely in $\Theta(n)$ sift-up operations.

Now, to each one sequence of priorities $s = (p_{1}, \ldots ,p_{n}) \in
Q_{n}$, we may associate a unique $n$-node tree $\alpha(s) = T_{n}$ in
the natural breadth-first, left-to-right order; by convention, we also let
$\alpha(\emptyset) = \Lambda$. In such a representation, $T_{n} = \alpha(s)$
is then an ordered $n$-node tree the ordering of which is the priority queue
(or heap) order, and it is thus built as an essentially complete binary heap
of size $n$. The correspondence $\alpha$ naturally represents the priority
queues $Q_{n}$ of size $n$ as ordered trees $T_{n}$ with $n$ nodes.

Conversely, to any ordered tree $T_{n}$ with $n$ nodes, we may associate a
binary tree with heap ordered nodes, that is an essentially complete binary
heap. Hence, there exists a correspondence $\beta$ mapping any given ordered
$n$-node tree $T_{n}$ onto a unique sequence of priorities $s = \beta(T_{n})
\in Q_{n}$; by convention we again let $\beta(\Lambda) = \emptyset$.

The correspondence is one-one, and it is easily seen that mappings $\alpha$
and $\beta$ are respective inverses.
\end{proof}

Let binary tournament trees represent each one of the above structures. Any
operation can thus be performed as if dealing with ordered trees $T_{n}$,
whereas binary tournament trees or permutations are really manipulated.
More precisely, since we know that $T_{n} \longleftrightarrow Q_{n}
\longleftrightarrow {\cal T}_{n} \longleftrightarrow S_{n}$, the cost of path
reversal performed on initial $n$-node trees $T_{n}$ which consist of a
root with $n-1$ children is {\em transported} from the $T_{n}$'s onto the
tournament trees $T \in {\cal T}_{n}$ and onto the permutations
$\sigma \in S_{n}$. In the following definitions (see Section 3.1 below),
we therefore let $\varphi(\sigma) \in S_{n}$ denote the ``reversed''
permutation which corresponds to the reversed tree $T_{n}$.
From this point the {\em first moment of the cost of path reversal},
$\varphi : T_{n} \rightarrow T_{n}$, can be derived, and a
straightforward proof technique of the result, distinct from the one in
section 3 below, is also detailed in the Appendix.

\section{Expected cost of path reversal, average message complexity of $\bm{\cal A}$}
It is fully detailed in the Introduction how the two data structures at hand 
are actually involved in algorithm $\cal A$ and the design of the algorithm 
takes place in~\cite{nta,tn}.

\subsection{Analysis}
Eq.~(13) proved in the Appendix, is actually sufficient to provide the average 
cost of path reversal. However, since we also desire to know the second moment 
of the cost, we do need the probability generating function of the probabilities 
$p_{n,k}$, defined as follows.

\medskip Let $h(T_{n})$ denote the height of $T_{n}$, {\em i.e.} the number 
of nodes on the path from the deepest node in $T_{n}$ to the root of $T_{n}$, 
and let $T \in {\cal T}_{n-1}$.
$$p_{n,k} \;= \;\Pr\{\mbox{cost of path reversal for}\ T_{n}\ \mbox{is}\ k\} %
\;=\; \Pr\{h (\varphi(T)) = k\}$$
is the probability that the tournament tree $\varphi(T)$ is of height $k$.
We also have
$$p_{n,k} \;=\; \Pr\{k\ \mbox{changes occur in the variable \textit{Last} 
of algorithm} {\cal A}\}.$$

More precisely, let a {\bf swap} be any interchanged pair of adjacent
prime cycles (see {\rm \cite[Vol.~3, pages 28-30]{kn}}) in a permutation
$\sigma$ of $[n-1]$ to obtain the ``reversed'' permutation
$\varphi_{x}(\sigma)$ corresponding to the path reversal performed at a node
$x \in T_{n}$, that is any interchange which occurs in the relative order of
the elements of $\varphi_{x}(\sigma)$ from the one of $\sigma$'s elements, and
let $N$ be the number of these swaps occurring from $\sigma \in S_{n-1}$ to
$\varphi_{x}(\sigma)$, then,
$$p_{n,k} \:=\; \frac{1}{(n-1)!}\, (\mbox{number of}\ \sigma \in S_{n-1}\ %
\mbox{for which}\ N=k),$$
since the cost of a path reversal at the root of an ordered tree such as
$T_{n}$ is zero.

\begin{lemma}
Let $P_{n}(z) = \sum_{k \geq 0} p_{n,k} z^{k}$ be the probability generating 
function of the $p_{n,k}$'s. We have the following identity,
$$P_{n}(z) \:=\; \prod_{j=1}^{n-1} \frac{z + j - 1}{j}\,.$$
\end{lemma}
\begin{proof}
We have $p_{1,0} = 1$ and $p_{1,k} = 0$ for all $k > 0$.
\par
A fundamental point in this derivation is that we are averaging not over
all tournament trees $T \in {\cal T}_{n-1}$, but {\em over all possible
orders} of the elements of $S_{n-1}$.
Thus, every permutation of $(n - 1)$ elements with $k$ swaps corresponds to
$(n - 2)$ permutations of $(n - 2)$ elements with $k$ swaps and one
permutation of $(n - 2)$ elements with $(k - 1)$ swaps. This leads directly
to the recurrence
$$(n-1)! p_{n,k} \;=\; (n-2)(n-2)! \,p_{n-1,k} \;+\; (n-2)! p_{n-1,k-1},$$
or
\begin{equation}
p_{n,k} =\; \left(1-\frac{1}{n-1} \right) p_{n-1,k} \;+\; %
\left(\frac{1}{n-1}\right) p_{n-1,k-1}.
\end{equation}
\par
Consider any permutation $\sigma = \langle \sigma_{1} \ldots \sigma_{n-1}
\rangle$ of $[n-1]$. Formula (1) can also be derived directly with the
argument that the probability of $N$ being equal to $k$ is
the simultaneous occurrence of $\sigma_{i} = j\; \; (1 \leq i,j \leq n-1)$ and
$N$ being equal to $k-1$ for the remaining elements of $\sigma$, {\em plus}
the simultaneous occurrence of $\sigma_{i} \neq j \; (1 \leq i,j \leq n-1)$ 
and $N$ being equal to $k$ for the remaining elements of $\sigma$. Therefore,
\begin{eqnarray*}
p_{n,k} & = & \Pr\{\sigma_{i} = j\} \times p_{n-1,k-1} \;+\; %
\Pr\{\sigma_{i} \neq j\} \times p_{n-1,k} \\
& = & \Big(1/(n-1)\Big) p_{n-1,k-1} \;+\; \Big(1-1/(n-1)\Big) p_{n-1,k}.
\end{eqnarray*}
\par
Using now the probability generating function $P_{n}(z) =\: \sum_{k \geq 0} p_{n,k} z^{k}$, 
we get after multiplying~(1) by $z^{k}$ and summing, 
$$(n - 1) P_{n}(z) \;=\; z P_{n-1}(z) \;+\; (n - 2) P_{n-1}(z),$$
which yields
\begin{eqnarray}
P_{n}(z) & = & \frac{z + n - 2}{n - 1} \; P_{n-1}(z) \nonumber \\
P_{1}(z) & = & z.
\end{eqnarray}

The latter recurrence~(2) telescopes immediately to

$$P_{n}(z) \; = \; \prod_{j=1}^{n-1} \frac{z + j - 1}{j} .$$
\end{proof}
\begin{remark}
The property proved by Trehel that the average number of
messages required by $\cal A$ is exactly the number of nodes at
height 2 in the reversed ordered trees $\varphi(T_{n})$ (see \cite{nta})
is hidden in the definition of the $p_{n,k}$'s. As a matter of fact, the
number of permutations of $[n]$ which contains exactly 2 prime cycles is
$\left[\begin{array}{c} n \\ 2\end{array}\right] 
\;=\; (n-1)!\, H_{n-1}$ (see~\cite{kn}), and whence the result.
\end{remark}
\begin{theorem}
The expected cost of path reversal and the average message complexity
of algorithm $\cal A$ is $\E(C_{n})\:=\; \overline{C_{n}} \:=\: H_{n-1}$, 
with variance $var(C_{n}) \:=\: H_{n-1} \;-\; H_{n-1}^{(2)}$.
Asymptotically, for large $n$,
$$\overline{C_{n}} \:=\; \ln n \;+\;\gamma \;+\; O(n^{-1})\ \quad \mbox{and}\
\quad var(C_{n}) \;=\; \ln n \;+\; \gamma \;-\; \pi^{2}/6 \;+\; O(n^{-1}).$$
\end{theorem}
\begin{proof}
By Lemma 3.1, the probability generating function $P_{n}(z)$ may be regarded
as the product of a number of very simple probability generating functions
(P.G.F.s), namely, for $1\leq j\leq n-1$,
$$P_{n}(z) =\; \prod_{1 \leq j \leq n-1} \Pi_{j}(z),\ \quad\ \mbox{with}\ %
\ \Pi_{j}(z) \;=\; \frac{j-1}{j} \;+\;\frac{z}{j}\,.$$

Therefore, we need only compute moments for the P.G.F. $\Pi_{j}(z)$, 
and then sum for $j = 1$ to $n - 1$. This is a classical property of P.G.F.s 
that one may transform products to sums.

\medskip \noindent Now, $\Pi_{j}'(1) \;=\; 1/j$ and $\Pi_{j}''(1) \:=\: 0$, 
and hence
$$\E(C_{n}) \;=\: \overline{C_{n}} \:=\; P_{n}'(1) \;=\; \sum_{j=1}^{n-1} %
\Pi_{j}'(1) \;= \: H_{n-1}.$$
Moreover, the variance of $C_{n}$ is
$$var(C_{n}) \:=\; P_{n}''(1) \;+\;P_{n}'(1) \;-\; P_{n}'^{2}(1),$$
and thus,
$$var(C_{n}) \:= \; \sum_{j=1}^{n-1} \frac{1}{j} \;-\; \sum_{j=1}^{n-1}
\frac{1}{j^{2}} \;=\; H_{n-1} \;-\; H_{n-1}^{(2)}.$$
\par
Since $H_{n-1}^{(2)} \;=\; \pi^{2}/6 \;-\; 1/n \:+\: O(n^{-2})$ when 
$n\rightarrow +\infty$, and by the asymptotic expansion of $H_{n}$, the
asymptotic values of $\overline{C_{n}}$ and of $var(C_{n})$ are easily obtained. 
(Recall that Euler's constant is $\gamma  = 0.57721\ldots$, thus 
$\gamma - \pi^{2}/6 \;=\: - 1.6772\ldots$)

Hence, $\overline{C_{n}} =\: .693\ldots \lg n \;+\; O(1)$, and
$var(C_{n}) =\: .693\ldots \lg n \;+\; O(1)$.
\end{proof}
Note also that, by a generalization of the central limit theorem to sums of
independent but nonidentical random variables, it follows that
$$\frac{C_{n} \:-\: {\overline C_{n}}}{(\ln n \,-\, 1.06772\ldots)^{1/2}}$$
converges to the normal distribution whe $n\rightarrow +\infty$.
\begin{proposition}
The worst-case message complexity of algorithm $\cal A$ is $O(n)$.
\end{proposition}
\begin{proof}
Let $\Delta$ be the {\em maximum} communication delay time in the
network and let $\Sigma$ be the {\em minimum} delay time for a process
to enter, proceed and release the critical section. \\
Set $q =\: \left\lceil \Delta/\Sigma \right\rceil$, the number of messages 
used in $\cal A$ is at most $(n-1) \:+\: (n-1)q \:=\: (n-1)(q+1) \:= O(n)$.
\end{proof}

\begin{remarks}
\item[1.]\ The one-to-one correspondence between ordered trees with $(n+1)$
nodes and the words of lenght $2n$ in the Dycklanguage with one type of
bracket is used in \cite{nta} to compute the average message complexity of
$\cal A$. Several properties and results connecting the depth of a Dyckword
and the height of the ordered $n$-node trees can be derived from the
one-to-one correspondences between combinatorial structures involved 
in the proof of Theorem 2.1.

\item[2.]\ In the first variant of algorithm $\cal A$ (see \cite{tn})
which is analysed here, a node never stores more than one request of some
other node and hence it only requires $O(\log n)$ bits to store the variables, 
and the message size is also $O(\log n)$ bits. This is not true of the second 
variant of algorithm $\cal A$ (designed in \cite{nt}). Though the constant 
factor within the order of magnitude of the average number of messages is claimed 
to be slightly improved (from 1 downto $.4$), the token now consists of a queue 
of processes requesting the critical section. 
Since at most $n-1$ processes belong to the requesting queue, the size of the 
token is $O(n\log n)$. Therefore, whereas the average message complexity 
is slightly improved (up to a constant factor), the message size increases 
from $O(\log n)$ bits to $O(n\log n)$ bits. The bit complexity is thus much
larger in the second variant~\cite{nt} of $\cal A$. Moreover, the state
information stored at each node is also $O(n\log n)$ bits in the second
variant, which again is much larger than in the first variant of $\cal A$.
\end{remarks}

\section{Waiting time and average waiting time of algorithm $\bm{\cal A}$}
Algorithm $\cal A$ is designed with the notion of {\em token}. Recall that a
node can enter its critical section only if it has the token. However, unlike
the concept of a token circulating continuously in the system, it is sent
from one node to another if and only if a request is made for it.
The token thus circulates strictly according
to the order in which the requests have been made. The queue is updated by
each node after executing its own critical section. The queue of requesting
processes is maintained at the node containing
the token and is transferred along with the token whenever the token is
transferred. The requesting nodes receive the token strictly according to the
order in the queue.

In order to simplify the analysis, the following is assumed.
\begin{itemize}
\item When a node is not in the critical section or is not already in the waiting
queue, it generates a request for the token at Poisson rate $\lambda$,
{\em i.e.} {\sl the arrival process is a Poisson process}.
\item Each node spends a constant time ($\sigma$ time units) in the critical
section, {\em i.e.} {\sl the rate of service is} $\mu = 1/\sigma$. Suppose we
would not assume a constant time spent by each process in the critical
section. $\sigma$ could then be regarded as the maximum time spent in the
critical section, since any node executes its critical section within a
finite time.
\item
The time for any message to travel from one node to any other node in the
complete network is {\sl constant} and is equal to $\delta$ (communication
delay). Since the message delay is finite, we assume here that every message
originated at any node is delivered to its destination in a {\sl bounded}
amount of time: in words, the network is assumed to be {\sl synchronous}
\end{itemize}
At any instant of time, a node has to be in one of the following two states:
\begin{enumerate}
\item {\sl Critical state.}

The node is waiting in the queue for the token or is
executing its critical section. In this state, it cannot generate any request
for the token, and thus the rate of generation of request for entering the
critical section by this node is zero.
\item
{\sl Noncritical state.}

The node is not waiting for the token and is not
executing the critical section. In this state, this node generates a request
for the token at Poisson rate $\lambda$.
\end{enumerate}

\subsection{Waiting time of algorithm $\bm{\cal A}$}
Let ${\cal S}_{k}$ denote the system state when exactly $k$ nodes are in the
waiting queue, including the one in the critical section, and let $P_{k}$
denote the probability that the system is in state ${\cal S}_{k}$, $0 \leq k
\leq n$.
In this state, only the remaining $n - k$ nodes can generate a request. Thus,
the net rate of request generation in such a situation is $(n - k) \lambda$.
Now the service rate is constant at $\mu$ as long as $k$ is positive,
{\em i.e.} as
long as at least one node is there to execute its critical section and the
service rate is $0$ when $k = 0$. The probability that during the time period
$(t,t+h)$ more than one change of state occur at any node is $o(h)$.
By using a simple birth-and-death process (see Feller, \cite{fe}), the
following theorem is obtained.

\begin{theorem}
When there are $k$ $(k > 0)$ nodes in the queue, the waiting time of a node
for the token is $w_{k} \:= \: (k - 1)(\sigma  +  \delta) \;+\; \sigma/2,$
where $\sigma$ denotes the time to execute the critical section and
$\delta$ the communication delay. \\
The worst-case waiting time is $w_{worst} \:\leq \: (n - 1)(\sigma + \delta) %
\;+\; \sigma/2 \;=\: O(n)$. \\
The exact expected waiting time of the algorithm is
$$\overline{w} \:=\: (\sigma + \delta) (\overline{n} - n P_{n}) \;-\;
(\delta + \sigma/2)(1  - P_{0} - P_{n}) \;+\; 2\delta P_{0},$$
where $\overline{n}$ denotes the average number of nodes in the queue and the
critical section.
\end{theorem}
\begin{proof}
Following equations hold,
\begin{eqnarray}
P_{0}(t + h) & = & P_{0}(t) (1 - n{\lambda}h) \;+\; P_{1}(t) %
\Big(1 \:-\: (n - 1){\lambda}h\Big) {\mu}h \;+\; o(h) \nonumber  \\
\frac{1}{h} \Big(P_{0}(t + h) - P_{0}(t)\Big) & = & - \,n{\lambda} P_{0}(t) %
\;+\; {\mu} P_{1}(t) \;-\; (n - 1){\lambda}{\mu}h P_{1}(t).
\end{eqnarray}

Under steady state equilibrium, $P_{0}'(t) = 0$ and hence,
\begin{equation}
\mu P_{1}(t) \; - \; n{\lambda} P_{0}(t) \: = \: 0, \; \; \mbox{and} \;
P_{1}(t) \; = \; n (\lambda/\mu) P_{0}(t).
\end{equation}

Similarly, for any $k$ such that $1 \leq k \leq n$, one can write
\begin{eqnarray*}
P_{k}(t + h) & = & P_{k}(t) \Big(1 \:-\: (n - k){\lambda}h\Big) (1 - {\mu}h) \\
& & +\ P_{k-1}(t) \Big((n - k + 1){\lambda}h\Big) (1 - {\mu}h) \\
& & +\ P_{k+1}(t) \Big(1 \:-\: (n - k - 1){\lambda}h\Big) ({\mu}h) \;+\; o(h),
\end{eqnarray*}
and
$$\frac{1}{h} \left(P_{k}(t + h) \:-\: P_{k}(t) \right) \; = \; %
- (n - k){\lambda} P_{k}(t) \;+\; (n - k + 1){\lambda} P_{k-1}(t) %
\;+\; {\mu} P_{k+1}(t) \;-\;{\mu} P_{k}(t).$$
\par
Classically, set $\rho = \lambda/\mu$. Proceeding similarly, since under
steady state equilibrium $P_{k}'(t) = 0$,
\begin{equation} 
P_{k+1}(t) \:=\; \left(1 + (n - k){\rho}\right) P_{k}(t) \:-\: (n - k + 1){\rho},
\end{equation}
and, for any $k$ ($1 \leq k \leq n$),
\begin{equation} \label{eq:pkdet}
P_{k}(t) \:=\; \frac{n!}{(n - k)!}\: \rho^{k} P_{0}(t) \;=\;
n^{\underline{k}} \rho^{k} P_{0}(t).
\end{equation}

For notational brevity, let $P_{k}$ denote $P_{k}(t)$. By Eq.~\ref{eq:pkdet}, 
we can now compute the average number of nodes in the queue and the critical 
section under the form
$$\overline{n} \: = \: \sum_{k=0}^{n} k P_{k} \: = \: P_{0} \sum_{k=0}^{n} k
n^{\underline{k}} \rho^{k},$$
since the system will always be in one of the $(n + 1)$ states 
${\cal S}_{0}, \ldots ,{\cal S}_{n}$, $\sum_{k} P_{k} = 1$.

Now using expressions of $P_{k}$ in terms of $P_{0}$ yields
$$\sum_{k=0}^{n} P_{k} \;=\; P_{0} \sum_{k=0}^{n} n^{\underline{k}} \rho^{k} \:=\: 1.$$
Thus,
\begin{equation}
P_{0} \:=\; \left(\sum_{k=0}^{n} n^{\underline{k}} \rho^{k} \right)^{-1}\ \quad %
\mbox{and}\ \quad P_{i} \:=\; \frac{n^{\underline{i}} \rho^{i}}{\left(\sum_{k=0}^{n} %
n^{\underline{k}}\rho^{k}\right)}.
\end{equation}

Let there be $k$ nodes in the system when a node $i$ generates a request.
Then $(k - 1)$ nodes execute their critical section, and one node executes
the remaining part of its critical section before $i$ gets the token. Thus,
when there are $k$ $(k > 0)$ nodes in the queue, the waiting time of a node
for the token is
\begin{center}
$w_{k} \:=\: (k - 1)$\ ({\sl time\ to\ execute\ the\ critical\ section}\ 
{\sl communication\ delay}\ $+$\ \\
$+$\ {\sl average\ remaining\ execution\ time}), or
\end{center}
\begin{equation}
w_{k} \:=\: (k - 1 ) (\sigma + \delta) \; + \; \sigma/2
\end{equation}
(where $\sigma$ denotes the time to execute the critical section and $\delta$
the communication delay).

When there are zero node in the queue, the waiting time is $w_{0} = 2\delta =$\  
{\sl total communication delay of one request and one token message}.
Hence the expected waiting time is
\begin{eqnarray*}
\E(w) & = & \overline{w} \:=\; \sum_{k=0}^{n-1} w_{k} P_{k} \;=\; %
2\delta P_{0} \;+\; \sum_{k=1}^{n-1} \left\{(k - 1)(\sigma  + \delta) %
\;+\; \frac{\sigma}{2}\right\} P_{k}, \\
& = & (\sigma  + \delta) \sum_{k=1}^{n-1} k P_{k} \;-\; (\sigma + \delta) %
\sum_{k=1}^{n-1} P_{k} \;+\; \sigma/2 \sum_{k=1}^{n-1} P_{k} \;+\; 2\delta P_{0},
\end{eqnarray*}
and
\begin{equation}
\overline{w} \:=\; (\sigma + \delta) (\overline{n} - n P_{n}) \;-\; %
(\delta + \sigma/2) (1 - P_{0} - P_{n}) \;+\; 2\delta P_{0}.
\end{equation}
\par
By Eq. (9), we know the exact value of the expected waiting time of a node
for the token in algorithm $\cal A$. Moreover,  by Eq. (8), the worst-case
waiting time is $w_{worst} \:\leq \; (n-1)(\sigma + \delta) \;+\; \sigma/2 \;=\; O(n).$
\end{proof}

\subsection{Average waiting time of algorithm $\bm{\cal A}$}
The above computation of $\overline{w}$ yields the asymptotic form of an
upper bound on the average waiting time $\overline{w}$ of $\cal A$.
\begin{theorem}
If $\rho < 1$, the average waiting time of a node for the token in $\cal A$
is asymptotically (when $n \rightarrow +\infty$),
$$\overline{w} \:\leq \: (\sigma + \delta) \left(n (1 - 2e^{-1/\rho}) \right) \;-\; %
(\delta + \sigma/2) (1 - e^{-1/\rho}) \;+\; O\left((\sigma/2 + 3\delta) e^{-1/\rho} %
\: \frac{\rho^{-n}}{n!}\right).$$
\end{theorem}
\begin{proof}
Assume $\rho = \lambda/\mu < 1$, when $n$ is large  we can bound
$\overline{w}$ from above in Eq.~(9) as follows,

First, $1 \:-\: P_{0} \:-\: P_{n} \;=\; 1 \:-\: P_{0} \:-\: n!\rho^{n} P_{0}$, 
where $P_{0} =\; \left(\sum_{i=0}^{n} n^{\underline{i}} \rho^{i} \right)^{-1}$.
\par
\noindent
Since $\rho < 1$,
$$P_{0} \:\leq \: \left(n! \rho^{n} e^{1/\rho} \right)^{-1},$$
and, for large $n$,
\begin{equation}
1 \:-\: P_{0} \:-\: P_{n} \;\leq \; 1 \:-\: \frac{1 + n! \rho^{n}}{n! \rho^{n} e^{1/\rho}} %
\:\leq \: 1 \;-\; \frac{1}{n! \rho^{n} e^{1/\rho}} \:-\: e^{\textstyle -1/\rho}.
\end{equation}

Next,
$$P_{n} = \; n! \rho^{n} P_{0} \;\geq\; \frac{\textstyle n! \rho^{n}}
{\textstyle n! \rho^{n} e^{1/\rho}} \;=\; e^{\textstyle - 1/\rho},$$
and
$$\overline{n} \;-\; n P_{n} \;=\; P_{0} \sum_{i=1}^{n} i
n^{\underline{i}} \rho^{i} \;-\; n P_{n}.$$
Now,
\begin{eqnarray*}
P_{0} \sum_{i=1}^{n} i n^{\underline{i}} \rho^{i} & = & \frac{\sum_{j=1}^{n}
(n - j) \rho^{-j}/j!}{\sum_{j=0}^{n} \rho^{-j}/j!}  \\
& = & n \;-\; \frac{n}{\sum_{j=0}^{n} \rho^{-j}/j!} \;-\; \frac{\sum_{j=1}^{n} j
\rho^{-j}/j!}{\sum_{j=0}^{n}  \rho^{-j}/j!}  \\
& \leq &  n \;-\; \frac{n}{e^{1/\rho}} \;-\; \frac{1}{n! \rho e^{1/\rho}}\,.
\end{eqnarray*}
Therefore,
\begin{equation}
\overline{n} \;-\; n P_{n} \; \leq \; n  \; - \;  n e^{-1/\rho} \;-\; %
\rho^{-1} e^{-1/\rho}/n! \;-\; n e^{-1/\rho} \; \leq \; n (1 \:-\: 2 e^{- 1/\rho}) %
\;- \;\rho^{-1} e^{-1/\rho}/n!
\end{equation}

Thus, by Eqs~(9), (10) and (11), we have
\begin{eqnarray*}
\overline{w}  & \leq  &  (\sigma + \delta) \left(n \left(1 \:-\: 2e^{-1/\rho} \right) %
\;-\; \frac{1}{n! \rho e^{1/\rho}}\right) \\
&  & \mbox{} \;-\; (\delta + \sigma/2) \left(1 \;-\; \frac{1}{n! \rho^{n} e^{1/\rho}} \;-\; %
e^{- 1/\rho} \right) \;+\; 2\delta \left( \sum_{i=0}^{n} n^{\underline{i}} \rho^{i} \right)^{-1},
\end{eqnarray*}
which yields the desired upper bound for $\rho < 1$. When $n$ is large,
$$\overline{w} \;\leq\; (\sigma + \delta) \left(n(1 - 2e^{-1/\rho}) \right)
\;-\; (\delta + \sigma/2) (1 - e^{-1/\rho}) \;+\;
O\left( (\sigma/2 + 3\delta)\: e^{-1/\rho} \frac{\rho^{-n}}{n!}\right).$$
\end{proof}
\begin{corollary}
If $\rho < 1$, and irrespective of the values of $\sigma$ and $\delta$, 
the worst-case waiting time of a node for the token in $\cal A$ is $O(n)$
when $n$ is large.
\end{corollary}

\section{Randomized bounds for the message complexity of $\bm{\cal A}$ in arbitrary networks}

The network considered now is general. The exact cost of a path reversal
in a rooted tree $T_{n}$ is therefore much more difficult to compute. The
message complexity of the variant $\cal A'$ of algorithm $\cal A$ for
arbitrary networks  is of course modified likewise.

Let $G = (V,E)$ denote the underlying graph of an arbitrary network, and
let  $d(x,y)$ the distance between two given vertices $x$ and $y$ of $V$.
The diameter of the graph $G$ is defined as $$D \: = \: \max_{x,y \in V}
d(x,y).$$
\begin{lemma}
Let $\left|V \right| = n$ be the number of nodes in $G$. The number of
messages $M_{n}$ used in the algorithm $\cal A'$ for arbitrary networks of
size $n$ is such that $0 \leq M_{n} \leq 2D$.
\end{lemma}
\begin{proof}
Each request for critical section is at least satisfied by sending zero
messages (whenever the request is made by the root), up to at most $2D$
messages, whenever the whole network must be traversed by the request
message.
\end{proof}

In order to bound $D$, we  make use of some results of B\'{e}la
Bollob\'{a}s about random graphs \cite{bo} which yield the following
summary results.

\subsection*{Summary results}
\begin{enumerate}
\item
Consider {\em very sparse} networks, {\em e.g.} for which the underlying
graph $G$ is just hardly connected. For {\em almost every} such network, the
worst-case message complexity of the algorithm is
$\Theta(\frac{\log n}{\log\log n})$.
\item
For {\em almost every} sparse networks ({\em e.g.} such that  $M = O(n/2)$),
the worst-case message complexity of the algorithm is $\Theta(\log n)$.
\item
For {\em almost every} $r$-regular network, the worst-case message
complexity of the algorithm is $O(\log n)$.
\end{enumerate}

\section{Conclusion and open problems}
The waiting time of algorithm $\cal A$ is of course highly dependent on the
values of many system parameters such as the service time $\sigma$ and the
communication delay $\delta$. Yet, the performance of $\cal A$, analysed 
in terms of average and worst-case complexity measures (time and message 
complexity), is {\em quite comparable} with the best existing distributed 
algorithms for mutual exclusion ({\em e.g.} \cite{ab,ra}). However, algorithm 
$\cal A$ is only designed for complete networks and is {\em a priori} not 
fault tolerant, although a fault tolerant version of $\cal A$ could easily 
be designed. The use of direct one-one correspondences between combinatorial 
structures and associated probability generating functions proves here a powerful 
tool to derive the expected and the second moment of the cost of path reversal; 
such combinatorial and analytic methods are more and more required to complete 
average-case analyses of distributed algorithms and data structures \cite{fl}. 
From such a point of view, the full analysis of algorithm $\cal A$ completed 
herein is quite general. Nevertheless, the simulation results obtained in the 
experimental tests performed in \cite{tn} also show a very good agreement with 
the average-case complexity value computed in the present paper. \par
Let $\cal C$ be the class of distributed tree-based algorithms for mutual
exclusion, {em e.g.} \cite{ab,ra}. There still remain  open questions about
$\cal A$. In particular, is the algorithm $\cal A$ average-case optimal
in the class $\cal C$~?
By the tight upper bound derived in \cite{gi}, we know that the amortized cost 
of path reversal is $O(\log n)$. It is therefore likely that the average complexity 
of $\cal A$ is $\Theta(\log n)$, and whence that algorithm $\cal A$  is average-case 
optimal in its class $\cal C$. The same argument can be derived from the fact that 
the average height of $n$-node binary search trees is $\Theta(\log n)$ \cite{fl}.

\section*{Appendix}
In Subsection 2.1, we developed the tools which make it possible to derive the
first two moments of the cost of path reversal, {\em viz.} the expected
cost and the variance of the cost (see Subsection 2.2). However, by Theorem~2.1, 
the average cost of path reversal may be directly proved to be $H_{n-1}$
by solving a straight and simple recurrent equation.
\begin{proposition}
The expected cost of path reversal and the average message complexity
of algorithm $\cal A$ evaluate to $\overline{\rm cost}(\varphi (T_{n}))
\:=\: \overline{C_{n}} \:=\: H_{n-1}$ (resp.).
\end{proposition}
\begin{proof}
Let $\overline{C_{k}}$\ be the average cost of path reversal performed
on an initial ordered $k$-node tree $T_{k}$, $1 \leq k \leq n$. $\overline{C_{k}}$\ 
is the {\em expected height} of $T_{k}$, or {\em the average number of
nodes on the path from the node} $x$ in $T_{k}$ at which the path reversal 
is performed {\em to the root of} $T_{k}$.

We thus have $\overline{C_{1}} + 1 =$\ average number of nodes on
the path from $x$ to the root in the tree $T_{1}$ with one node
($\overline{C_{1}} + 1 = 1$), $\overline{C_{2}} + 1 =$\ average number
of nodes on the path from $x$ to the root in a tree $T_{2}$ with two nodes
($\overline{C_{2}} + 1 = 2$), etc. And therefore, the following identity
holds
\begin{equation}
\overline{C_{k+1}} \:=\; \frac{1}{k} \left((\overline{C_{1}} + 1) \:+\: (\overline{C_{2}} + 1) %
\:+\cdots +\: (\overline{C_{k}} + 1) \right)\quad \mbox{for}\ k = 1, \ldots, n.
\end{equation}

We can rewrite this recurrence in two equivalent forms:
\begin{eqnarray*}
k \overline{C_{k+1}} & = & k \:+\: \overline{C_{1}} \:+\: %
\overline{C_{2}} \:+\cdots +\: \overline{C_{k}}\ \qquad (k\ge 1) \\
(k - 1) \overline{C_{k}} & = & (k - 1) \: + \: \overline{C_{1}} \:+\:
\overline{C_{2}}\: + \cdots + \: \overline{C_{k-1}}\ \qquad (k\ge 2).
\end{eqnarray*}
Substracting these equations yields
$$k \overline{C_{k+1}} \:=\: 1 \:+\: k \overline{C_{k}}\ \ (k\ge 2)\ %
\quad \mbox{and}\ \ \overline{C_{1}}=0.$$

\noindent Hence, 
\begin{eqnarray*}
\overline{C_{k+1}} & = & \overline{C_{k}} \; + \; 1/k,\quad (k \geq 2) \\
\overline{C_{1}} & = & 0,
\end{eqnarray*}
and the general formula $\overline{C_{k+1}} \:=\: H_{k}$ follows.
Setting $k = n$ finally gives the average cost of path reversal for ordered 
$n$-node trees,
\begin{equation}
\overline{C_{n}} \:=\: H_{n-1}.
\end{equation}
\end{proof}

Note that if we let $\overline{\left|LB(T)\right|} = %
\overline{\left|RB(T)\right|}$\ denote the mean length of a right or 
left branch of a tree $T \in {\cal T}_{n-1}$, we also have (by Lemma 2.1)
$$\overline{\rm cost}(\varphi(T_{n})) \:=\: \overline{\left|LB(T)\right|},$$
where $\overline{\left|LB(T)\right|}$\ is averaged over all the $(n - 1)!$\ 
binary tournament trees in ${\cal T}_{n-1}$.

\end{document}